\documentclass[epj-spec]{svjour}
\usepackage{amssymb,graphicx,xspace,color,textcomp}
\newcommand{\tc}{$T_c$\xspace}

\newcommand{\tco}{\ensuremath{T_{CO}}\xspace}

\newcommand{\tlt}{\ensuremath{T_{LT}}\xspace}
\newcommand{\tht}{\ensuremath{T_{HT}}\xspace}

\newcommand{\lasr}{\ensuremath{\mathrm{La_{1.85}Sr_{0.15}CuO_4}}\xspace}
\newcommand{\lasrx}{\ensuremath{\mathrm{La_{2-x}Sr_xCuO_{4}}}\xspace}
\newcommand{\labax}{\ensuremath{\mathrm{La_{2-x}Ba_xCuO_{4}}}\xspace}

\newcommand{\landx}{\ensuremath{\mathrm{La_{1.6-x}Nd_{0.4}Sr_xCuO_{4}}}\xspace}

\newcommand{\lasrndx}{\ensuremath{\mathrm{La_{2-x-y}Sr_xNd_yCuO_{4}}}\xspace}

\newcommand{\lesco}{\ensuremath{\mathrm{La_{1.8-x}Eu_{0.2}Sr_xCuO_4}}\xspace}

\newcommand{\lasrre}{\ensuremath{\mathrm{La_{2-x-y}RE_ySr_xCuO_{4}}}\xspace}

\begin{document}
\title{Nernst Effect of stripe ordering La$_{1.8-x}$Eu$_{0.2}$Sr$_x$CuO$_4$}
\author{Christian Hess\inst{1}\fnmsep\thanks{\email{c.hess@ifw-dresden.de}} \and Emad M. Ahmed\inst{1}\fnmsep\thanks{Present address: Solid State Physics Department, National Research Center, Postal Code 12622, Dokki, Cairo, Egypt} \and Udo Ammerahl\inst{2} \and Alexandre Revcolevschi\inst{2} \and Bernd B\"uchner\inst{1}}
\institute{IFW-Dresden, Institute for Solid State Research, P.O. Box 270116, D-01171 Dresden, Germany \and Laboratoire de Physico-Chimie de L'Etat Solide, ICMMO, UMR 8182, Universit\'{e} Paris-Sud, 91405 Orsay, Cedex, France}
\abstract{We investigate the transport properties of \lesco ($x=0.04$, 0.08, 0.125, 0.15, 0.2) with a special focus on the Nernst effect in the normal state. Various anomalous features are present in the data. For $x=0.125$ and 0.15 a kink-like anomaly is present in the vicinity of the onset of charge stripe order in the LTT phase, suggestive of enhanced positive quasiparticle Nernst response in the stripe ordered phase. At higher temperature, all doping levels except $x=0.2$ exhibit a further kink anomaly in the LTO phase which cannot unambiguously be related to stripe order. Moreover, a direct comparison between the Nernst coefficients of stripe ordering \lesco and superconducting \lasrx at the doping levels $x=0.125$ and $x=0.15$ reveals only weak differences. Our findings make high demands on any scenario interpreting the Nernst response in hole-doped cuprates.} 
\maketitle
\newpage
\section{Introduction}
\label{intro}
The Nernst effect of unconventional superconductors has recently attracted a lot of attention for several reasons \cite{Hagen1990,Ri1994,Xu2000,Wang2001,Wang2006,Cyr-Choiniere2009,Hackl2009a,Hackl2009,Behnia2009,Daou2010,Hackl2010,Kondrat2010}. In general, for type-II superconductors, the Nernst response is strongly enhanced by the movement of magnetic flux lines (vortices) \cite{Hagen1990,Ri1994,Huebener1969,Otter1966}. Based on this very fundamental property, the unusual enhancement of the Nernst coefficient in the normal state of the cuprates at temperatures much higher than the critical temperature $T_c$ has been interpreted as the signature of vortex fluctuations \cite{Xu2000,Wang2001,Wang2006}. More specifically, it was proposed that in the pseudogap phase above $T_c$ long-range phase coherence of the superconducting order parameter is lost while the pair amplitude remains finite. This interpretation of the data has been challenged only recently and it has been proposed that Fermi surface distortions due to stripe or spin density wave (SDW) order could lead to an enhanced Nernst effect in the cuprates \cite{Cyr-Choiniere2009,Hackl2009a,Hackl2010}. In particular, for stripe ordering \lesco and \landx, an enhanced positive Nernst signal at elevated temperature has been associated with a Fermi surface reconstruction due to stripe order \cite{Cyr-Choiniere2009}.
Very recently, a strong anisotropy of the Nernst coefficient arising from the broken rotation symmetry of electron-nematic order has been discussed both experimentally and theoretically \cite{Daou2010,Hackl2009}.

The tendency towards the segregation of spins and holes in cuprate superconductors is much under debate with respect to the nature of superconductivity and the pseudogap phase \cite{Cyr-Choiniere2009,Daou2010,Tranquada2004,Hinkov2004,Hinkov2007,Kohsaka2007,Hinkov2008}. Static stripe order, i.e. stripe arrangements of alternating hole-rich and antiferromagnetic regions has been observed in the prototype system \labax \cite{Tranquada2004,Fujita2004,Huecker2010,Tranquada2008,Wen2008a,Xu2007} and the closely related compounds \lesco and \landx \cite{Fink2009a,Tranquada95}. In these materials an intimate interplay between structure, stripe order and superconductivity is present. More specifically, bulk superconductivity is suppressed in favor of static stripe order where the latter is stabilized through a particular tilting pattern of the $\rm CuO_6$ octahedra in the low-temperature tetragonal structural phase (LTT-phase) \cite{Tranquada2008,Tranquada95,Buechner1991,Buechner1993,Buchner94,Klauss00}. 

In \lesco, the LTT phase is present at lowest temperature over a wide doping range from $x=0$ to $x=0.2$, which is in contrast with \labax where the LTT phase is only present in a limited doping range around $x=1/8$ \cite{Axe1989}. In addition, irrespective of doping, the transition temperature extends up to rather high temperatures $T_{LT}\approx 120\pm10$~K, i.e. much higher than in  \lasrndx where $T_{LT}\approx 70$~K \cite{Buechner1991,Buechner1993,Tranquada96a}.
The suppression of bulk superconductivity extends in \lesco over a wide doping range up to $x\lesssim0.2$. Around $x=0.2$  the tilt angle of the octahedra and hence the buckling of the plane which decreases with increasing hole doping become smaller than a critical value \cite{Buchner94,Klauss00}. At $T>T_{LT}$ the structure enters the low temperature orthorhombic (LTO) phase in which the buckling pattern of the $\rm CuO_2$ planes does not support static stripe order \cite{Klauss00}.\footnote{Note that the LTO phase is present in \lasrx down to lowest temperature \cite{Birgeneau87,Braden1994}.}  At even higher temperatures the structure enters a further tetragonal phase (so-called high temperature tetragonal phase, HTT) at $T_{HT}$. In the case of \lesco, $T_{HT}>300$~K, at $x\leq0.15$ and $T_{HT}\approx220$~K for $x=0.2$ \cite{Klauss00}.

In this paper we investigate the transport properties of \lesco ($x=0.04$, 0.08, 0.125, 0.15, 0.2) with a special focus on the Nernst effect in the normal state. Various anomalies are observed in the data. For $x=0.125$ and 0.15 a kink-like anomaly is present in the LTT phase at a similar temperature as the onset of charge stripe order as seen in diffraction experiments \cite{Fink2009a}. At higher temperature, all doping levels except $x=0.2$ exhibit a further kink anomaly in the LTO phase which cannot unambiguously be related to stripe order, in contrast with previous statements \cite{Cyr-Choiniere2009}. Moreover, a direct comparison between the Nernst coefficients of stripe ordering \lesco and superconducting \lasrx at the doping levels $x=0.125$ and $x=0.15$ reveals only weak differences. The latter finding makes high demands on any scenario for interpreting the Nernst response in hole-doped cuprates. 

\section{Experimental}
We have prepared single crystals of \lesco ($x=0.04$, 0.08, 0.125, 0.15, 0.2) and of \lasr utilizing the traveling solvent floating zone technique. The crystals were investigated previously by means of magnetic susceptibility \cite{Huecker2004}, nuclear magnetic resonance (NMR) \cite{Simovic2003}, thermal transport \cite{Hess2003b}, resonant soft x-ray scattering (RSXS) \cite{Fink2009a} and neutron scattering \cite{Huecker2007}.
Measurements of the in-plane Nernst, Seebeck and Hall coefficients and of the electrical resistivity were performed on cube-shaped pieces cut along the principal axes with a typical length of 2~mm along the measuring direction. The Seebeck and Nernst coefficients were measured using a home-made heater and an Au-Chromel differential thermocouple for determining the temperature gradient on the sample \cite{Kondrat2010,Hess2003b}. The Nernst coefficient was determined as a function of temperature $T$ at constant magnetic field $B=8~$T with two opposite field polarizations, to compensate the longitudinal thermal voltages. The Hall effect was measured using a standard four-point method by sweeping the magnetic field at stabilized temperatures.

\section{Results}
\subsection{Resistivity}
\begin{figure}[t]
\includegraphics[clip,height=1\columnwidth,angle=270]{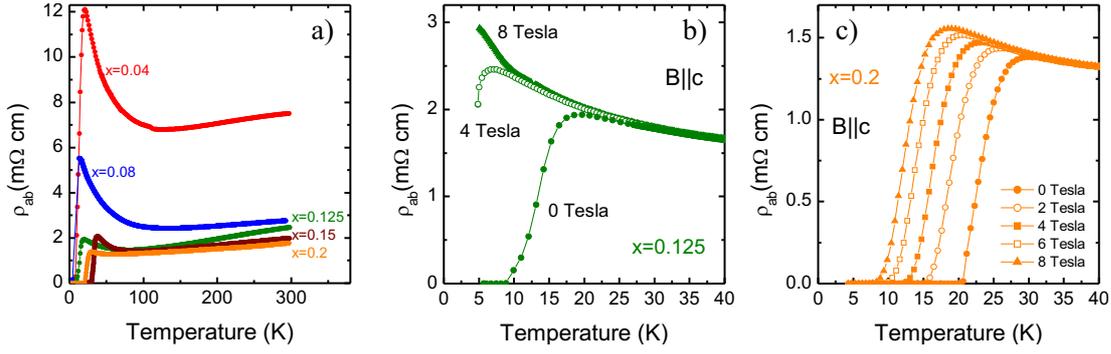}
\caption{a) In-plane resistivity $\rho_{ab}$ of \lesco with $x=0.04$, 0.08, 0.125, 0.15, 0.2 as a function of temperature. b) and c) $\rho_{ab}$ for various magnetic fields $B||c$ for $x=0.125$  and $x=0.2$, respectively.} 
\label{Fig_rho}
\end{figure}

Figure~\ref{Fig_rho} presents our results for the in-plane resistivity $\rho_{ab}$. 
The resistivity (Figure~\ref{Fig_rho}a) exhibits at room temperature a systematic decrease with increasing doping level, consistent with previous results \cite{Takagi1992,Hucker1998,Noda1999,Ichikawa2000}. 
Note, that the absolute value of $\rho_{ab}$ of our crystals is lower than that of previously published data on polycrystalline samples \cite{Hucker1998} but somewhat higher than that of \lasrndx crystals \cite{Noda1999,Ichikawa2000} or \labax \cite{Li2007,Tranquada2008} which might indicate an intrinsically enhanced carrier scattering in the \lesco material. 
The critical temperatures of superconductivity \tc are strongly reduced with respect to those of \lasrx at the respective doping levels and, moreover, for $x<0.2$, superconductivity can easily be suppressed by a laboratory magnetic field (see Figures~\ref{Fig_rho}b and \ref{Fig_rho}c for representative examples). 
Also consistent with previous findings, all curves exhibit a low-temperature upturn in the LTT phase, characteristic of carrier localization, prior to the onset of superconductivity. Note that the size of the upturn is non-monotonic and shows a minimum at $x=1/8$ \cite{Hucker1998}. Furthermore it is worth mentioning that, except for $x=0.04$, no anomaly is present at $T_{LT}\approx 125$~K \cite{Klauss00} which indicates that the carrier localization due to stripe order sets in smoothly at $T<T_{LT}$ which is in contrast with the abrupt change of $\rho(T)$ that is observed in \landx where $T_{LT}$ is much lower ($\sim70$~K) \cite{Ichikawa2000}.

\subsection{Hall effect}

The in-plane Hall coefficient $R_H$ shown in Figure~\ref{Fig_RH_S}a shows a similarly systematic doping evolution as the resistivity. In particular, at room temperature, $R_H$ decreases monotonically with increasing doping level. Upon lowering the temperature, $R_H$ of all samples, except at $x=0.2$ (which exhibits a rather conventional $R_H$), becomes strongly temperature dependent. To be specific, for $x=0.08$, 0.125 and 0.15, $R_H(T)$ increases slightly with decreasing temperature $T$ at constant slope until a characteristic temperature is reached below which $R_H$ steeply decreases. A similar drop of the Hall coefficient has been reported in \lasrndx, with the difference that in that material $R_H$ drops in a jump-like manner at $T_{LT}$ and then trends to even smaller values as  temperature decreases \cite{Nakamura92,Noda1999}. This peculiar behavior has been attributed to the abrupt onset of stripe order at \tlt \cite{Noda1999}. Our data for \lesco thus suggest that the characteristic temperature below which stripe order affects $R_H$ is clearly decoupled from the $\mathrm{LTO}\rightarrow\mathrm{LTT}$ transition which occurs at a much higher $T_{LT}\approx120\pm10$~K.
For \lesco with $x=0.125$ and $x= 0.15$ a direct comparison of this characteristic temperature with the stripe charge ordering temperature \tco as seen in RSXS measurements \cite{Fink2009a} is possible.
Interestingly, this comparison reveals that for $x=0.125$ the characteristic temperature is identical with $T_{CO}=80$~K. For $x=0.15$, however, the drop of $R_H$ occurs at about 50~K, i.e., clearly at a lower temperature than $T_{CO}=65$~K. This finding highlights the special importance of $x=1/8$ and suggests that the impact of stripe order on the electronic properties is less complete once the hole content deviates from this value.

\subsection{Thermopower}
\begin{figure}[t]
\includegraphics[clip,height=1\columnwidth,angle=270]{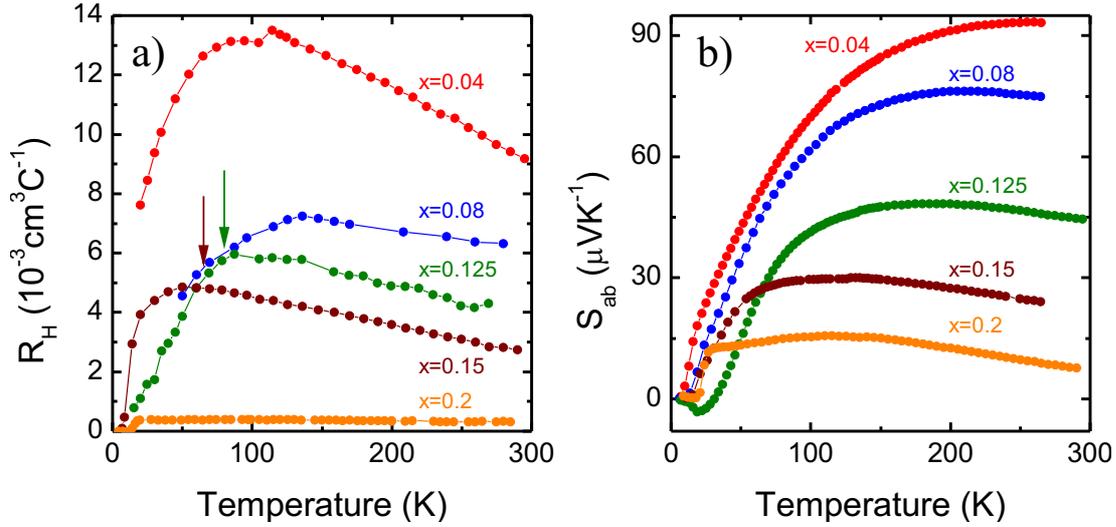}
\caption{a) In-plane thermopower $S$ (a) and in-plane Hall coefficient $R_H$ (b) of \lesco with $x=0.04$, 0.08, 0.125, 0.15, 0.2 as a function of temperature. Arrows mark the charge stripe ordering temperatures \tco for $x=0.125$, 0.15 as seen in RSXS experiments \cite{Fink2009a}.} 
\label{Fig_RH_S}
\end{figure}

The in-plane thermopower $S$ of our samples is displayed in Figure~\ref{Fig_RH_S}b. The systematic doping evolution agrees well with previous results on poly- and single crystals \cite{Hucker1998,Nakamura92}. No obvious anomalies that could be related to the onset of stripe order are discernible in the data. Note, however, that $S\geq0$ for all samples except for the low-$T$ behavior at $x=0.125$ doping, where $S\leq0$ below $\sim30$~K, which is a generic feature of stripe order around 1/8 doping \cite{Hucker1998,Nakamura92,Chang2010}.

\begin{figure}[ht]
\includegraphics[clip,width=0.97\columnwidth]{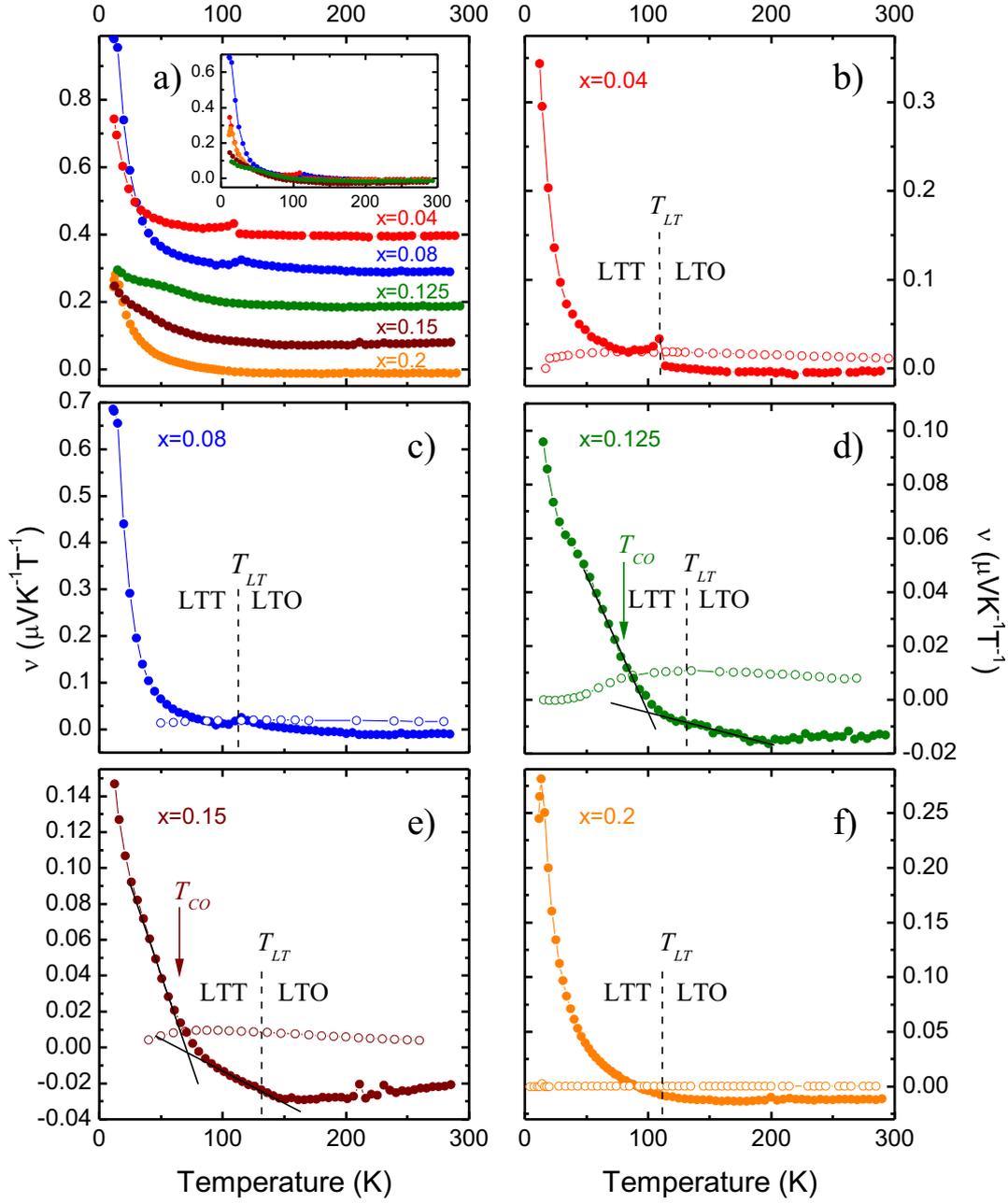}
\caption{Nernst coefficient $\nu$ of \lesco ($x=0.04$, 0.08, 0.125, 0.15, 0.2) as a function of temperature. a) Overview on all data. The presented curves have been shifted for clarity. Inset: all curves at the same scale. b)-f) Temperature dependence of $\nu$ (full symbols) and of $S\tan\theta /B$ (open symbols) for each doping level. Solid lines are linear extrapolations of $\nu(T)$ in order to extract $T_{\nu*}$. Arrows mark the charge stripe ordering temperatures \tco for $x=0.125$, 0.15 as seen in RSXS experiments \cite{Fink2009a}.} 
\label{Fig_all_nernst}
\end{figure}

\subsection{Nernst effect}

Figure~\ref{Fig_all_nernst} presents the temperature dependence of the Nernst coefficient $\nu=\frac{E_y}{\nabla TB}$ (with $E_y$ the transverse electrical field and $B$ the magnetic field) of all investigated samples. As can be inferred from panel a) of the figure, the overall magnitude of the Nernst coefficient is very similar for temperatures that can clearly be attributed to the normal state, i.e. at $T\gtrsim50$~K. In this temperature range, relatively small anomalies are present in the $T$-dependence of the Nernst coefficient which will be discussed in more detail further below. At lower temperatures ($T\lesssim50$) all curves exhibit a strong low-temperature rise, the magnitude of which, however, develops very non-monotonically as a function of doping with a clear minimum at $x=1/8$. We hence attribute this low-temperature rise in the Nernst coefficient to fluctuations of the superconducting order parameter which experiences a severe suppression in the presence of stripe order \cite{Klauss00}, being strongest at $x=1/8$.

The panels b) to f) of Figure~\ref{Fig_all_nernst} highlight the temperature dependence of the Nernst coefficient for each doping level. In addition to the actual Nernst coefficient $\nu$ (full symbols), we also display the quantity $S\tan\theta /B$ (open symbols) which is related to the Nernst coefficient via the expression \cite{Wang2001,Sondheimer1948}
\begin{equation}
{\nu=(\frac{\alpha_{xy}}{\sigma}-S\tan \theta)\frac{1}{B}}. \label{sondheimer}
\end{equation}
Here  $\tan \theta$  is the Hall angle, $\sigma$ the electrical conductivity, and $\alpha_{xy}$ the off-diagonal Peltier conductivity. Since $S\tan\theta /B$  can be easily computed from our data for $\rho$, $R_H$ and $S$, equation~\ref{sondheimer} allows to judge whether any of the observed anomalies is related to  $\alpha_{xy}$, i.e. a true off-diagonal thermoelectric quantity or to an anomalous behavior in the complementary transport coefficients.

We first examine the data with respect to any impact of the structural transition at \tlt which is present in all compounds \cite{Klauss00,Hess2003b}. As is obvious in the data, a jump-like anomaly is present at \tlt for $x=0.04$ and $x=0.08$, where a larger jump size is found for the lower hole content. At higher doping levels no anomaly is observed. These findings suggest that the Nernst response couples directly to structural distortions of the $\rm CuO_2$-plane. Note, that $S\tan\theta /B$ does not contribute significantly to the observed jumps. The decrease of the jump size (towards its complete disappearance at $x\geq0.125$) with increasing Sr doping level is consistent with a concomitant decreasing tilting angle of the $\rm CuO_6$ octahedra, i.e. a decreasing buckling of the  $\rm CuO_2$ plane \cite{Buchner94}. However, the rapid decrease of the anomaly with doping suggests that not only structural (degree of buckling) but also electronic details (hole content) play a decisive role in this regard.

For $x=0.125$, a further inspection of the data reveals two kink-like features which deserve closer consideration (see figure~\ref{Fig_all_nernst}d). One occurs deep in the LTT phase at $T_{\nu*}\approx100$~K and marks a strong change of slope, the other occurs in the LTO phase at $T_\nu\approx180$~K. 
At first glance, none of these two anomalies is in an obvious way related to the onset temperature of neither the charge stripe nor the spin stripe order which are known as $T_{CO}=80$~K \cite{Fink2009a} and $T_{SO}\approx45$~K \cite{Klauss00,Huecker2007}, respectively, i.e. at clearly much lower temperature than both $T_{\nu*}$ and $T_{\nu}$. However, the kink at $T_{\nu*}$ and the ordering temperature for charge stripes \tco, detected by RSXS experiments, occur at a not too  different temperature, which may indicate a close connection between both. The diffraction experiment requires a certain correlation length of the stripe order to be exceeded in order to generate a superlattice reflection. It is therefore thinkable that short range stripe order already develops at $T_{\nu*}$, giving rise to an enhanced Nernst coefficient at this temperature, which eventually becomes long range at \tco as seen in RSXS. In contrast to this,  the kink-temperature $T_{\nu}$ seems to be too high to account for the stripe ordering phenomena in the LTT phase in an obvious manner.

We find a very similar situation also at $x=0.15$. Here, we observe $T_{\nu*}\approx70$~K and $T_{\nu}\approx145$~K with respect to $T_{CO}=65$~K \cite{Fink2009a} and $T_{SO}\approx45$~K \cite{Klauss00,Huecker2007}. Note that $T_{\nu*}\approx T_{CO}$ in this case which corroborates the above conjecture that the kink at $T_{\nu*}$ could be related to the formation of charge stripe order. 

\begin{figure}[t]
\includegraphics[clip,width=0.97\columnwidth]{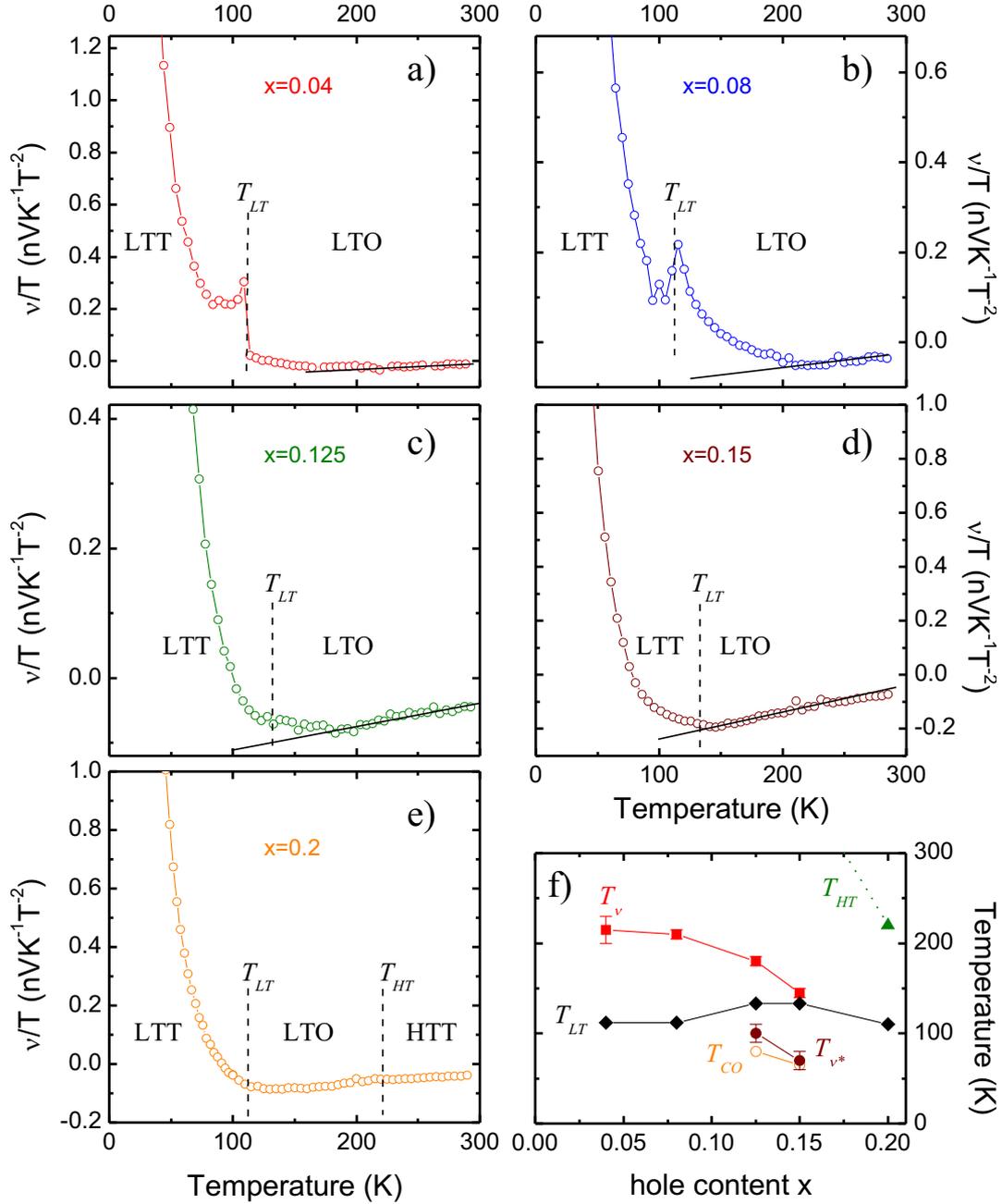}
\caption{a)-e) $\nu/T$ of \lesco ($x=0.04$, 0.08, 0.125, 0.15, 0.2) as a function of temperature. Solid lines are linear extrapolations of the high temperature linear behavior of $\nu(T)$ in order to extract $T_{\nu}$. f) Phase diagram showing \tlt ($\blacklozenge$), \tht ($\blacktriangle$), $T_{\nu}$ ($\blacksquare$) $T_{\nu*}$ (\textbullet) and \tco (\textopenbullet) from RSXS measurements \cite{Fink2009a}.} 
\label{Fig_nut}
\end{figure}
The comparison of these findings with other doping levels reveals that similarly salient features are not present. Cyr-Choiniere et al. have recently suggested to plot the quantity $\nu/T$ versus temperature as a useful representation of the Nernst effect to detect a potentially present anomalous enhancement of the Nernst coefficient due to  stripe order. The proposed criterion for detecting an enhanced Nernst response is based on the assumption that $\nu/T\propto T$ when no Fermi surface reconstruction and no contribution from superconductivity are present \cite{Cyr-Choiniere2009}.
Panels a) to e) of Figure~\ref{Fig_nut} display this representation for our data. As can be seen in the figure, at all doping levels up to $x=0.15$, $\nu/T$ is indeed linear at high $T$ and deviates from this linearity at a characteristic temperature. Interestingly, this characteristic temperature decreases monotonically upon increasing doping, and, for $x=0.125$ and 0.15 we find this characteristic temperature being identical to that of the high-temperature kink at $T_{\nu}$. We stress that here we did not consider the doping level $x=0.2$ despite  $\nu/T$ being also linear in $T$ at high temperature since this sample undergoes two structural transitions in the region of interest. One is the  $\rm LTO\rightarrow LTT$ transition at $T_{LT}\approx110$~K, the other is the  $\rm HTT\rightarrow LTO$ transition at $T_{HT}\approx220$~K.

\section{Discussion}
We summarize our findings in the phase diagram shown in Figure~\ref{Fig_nut}f. The main experimental finding is the rather good agreement between the lower kink temperature $T_{\nu*}$ and the charge stripe ordering temperature \tco for those doping levels where the stripe order has been experimentally detected by diffraction experiments. Qualitatively, the enhanced $\nu$ at $T<T_{CO}$ seems consistent with theoretical results by Hackl et al., who calculated the temperature dependence of the quasiparticle Nernst effect for $p=1/8$ stripe order in a mean field approach \cite{Hackl2010}.

The origin of the high temperature anomaly at $T_\nu$ remains unclear. In a similar analysis Cyr-Choiniere et al. have interpreted $T_\nu$ as the onset temperature of Fermi reconstruction due to stripe order \cite{Cyr-Choiniere2009}.\footnote{It seems noteworthy at this point that Cyr-Choiniere et al. determined for $x=0.125$ a significantly lower $T_\nu\approx140$~K and did not observe a kink at $T_{\nu*}$ despite an overall good agreement of our data with their result \cite{Cyr-Choiniere2009}.} Our data do not support this interpretation since clear anomalies corresponding to the true stripe ordering in the LTT phase are found at $T_{\nu*}$. One might speculate though that the apparently weakly enhanced Nernst response at $T\leq T_\nu$ is related to a stripe fluctuation which could extend up to these elevated temperatures \cite{Hackl2010,Xu2007}. On the other hand, one cannot exclude that subtle structural effects unrelated to electronic order are the actual cause of the slight enhancement at $T_\nu$. For example, soft phonon type precursors of the $\rm LTO\rightarrow LTT$ transition are known to be ubiquitous in the LTO phase of both \lasrre (RE=Rare Earth), which undergoes the $\rm LTO\rightarrow LTT$ transition, and \lasrx, which, remains in the LTO phase down to lowest temperature \cite{Pintschovius90,Martinez91,Keimer93}. This conjecture is supported by the jump-like response of the Nernst coefficient at \tlt for the lower doping levels.

\begin{figure}[t]
\includegraphics[clip,width=1\columnwidth]{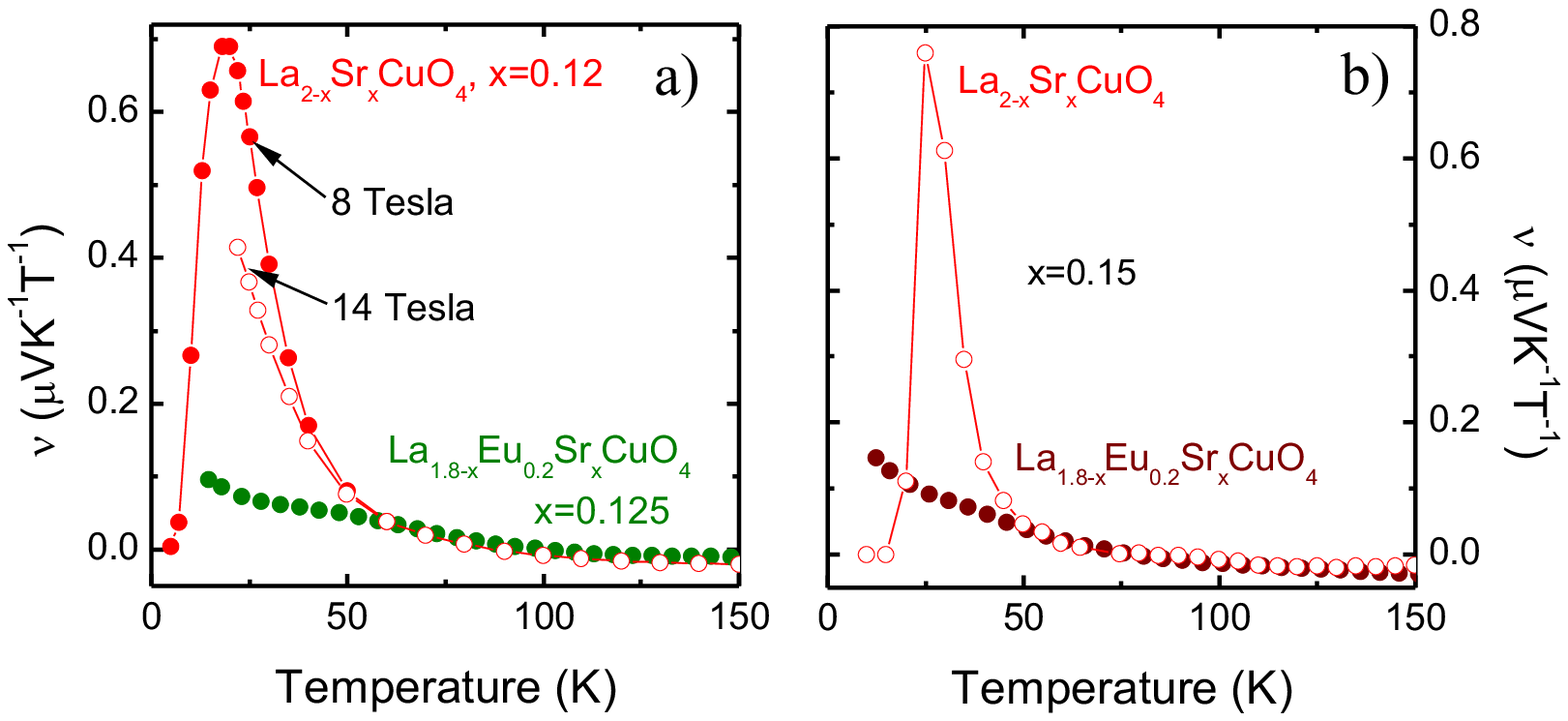}
\caption{Nernst coefficient $\nu$ of \lesco compared to that of \lasrx a) \lesco at $x=0.125$ and \lasrx at $0.12$ (data taken from Ref.~\cite{Wang2006}). b) \lesco and \lasrx at $x=0.15$.} 
\label{Fig_LESCO_LSCO}
\end{figure}
It is interesting to point out that the afore discussed salient features attributed to stripe order are, in fact, very subtle anomalies. To illustrate this, we compare in Figure~\ref{Fig_LESCO_LSCO} the Nernst coefficient of both stripe ordering \lesco and non-stripe ordering, bulk superconducting \lasrx for the doping levels $x\approx0.125$ and $x=0.15$. As can be seen in the figure, the Nernst-effect of both variants is very similar in the normal state, i.e. at $T\gtrsim50$~K. This makes high demands on any scenario for interpreting the Nernst response in hole-doped cuprates. Considering the stripe ordering scenario, these data suggest that the magnitude of the Nernst response for static and fluctuating stripes is practically the same (apart from the subtle anomalies at $T_\nu$). Theoretical treatments for the Nernst response in the presence of fluctuating stripes are thus required. On the other hand, a vortex fluctuation scenario which attributes the enhanced Nernst coefficient in the normal state largely to superconducting fluctuations should explain why in stripe ordering phases the presence of static stripes and the suppression of bulk superconductivity at low $T$ has only a weak influence on the normal state Nernst effect.

\section{Summary}
We have investigated the transport properties of \lesco ($x=0.04$, 0.08, 0.125, 0.15, 0.2) with a special focus on the Nernst effect in the normal state. For $x=0.125$ and 0.15 a kink-like anomaly is present in the LTT phase which roughly coincides with the onset of charge stripe order, suggestive of an enhanced positive Nernst response in the stripe ordered phase. At higher temperature, all doping levels except $x=0.2$ exhibit a further kink anomaly in the LTO phase with unclear origin. Possible explanations could involve an enhanced Nernst effect due to stripe fluctuations or due to structural fluctuations. A direct comparison between the Nernst coefficients of stripe ordering \lesco and superconducting \lasrx at the doping levels $x=0.125$ and $x=0.15$, reveals only weak differences in the normal state. 

\section{Acknowledgements}
We are grateful to the Deutsche Forschungsgemeinschaft for supporting our research work through the Research Unit FOR538 (Grant No.BU887/4). Furthermore we thank Markus H\"ucker for inspiring discussions.

%
%
%

%
%


\end{document}